# Energetics of star–disc encounters in the non-linear regime

S.M. Hall,[1] C.J. Clarke[2] and J.E. Pringle[1]
[1] *Institute of Astronomy, Madingley Road, Cambridge, CB3 0HA*
[2] *School of Mathematical Sciences, Queen Mary & Westfield College, Mile End Road, London, E1 4NS*

**ABSTRACT**

We investigate the response of a circumstellar accretion disc to the fly-by of a perturbing mass on a parabolic orbit. The energy and angular momentum transferred during the encounter are calculated using a reduced three-body method. In almost all close encounters the energy and angular momentum transfer is dominated by disc material becoming unbound from the system, with the contributions from close disc particle – star encounters being significant. For more distant encounters with some prograde element to the motion the disc material loses energy and angular momentum to the perturber's orbit through a resonance feature. The magnitude of the energy transfer calculated in our simulations is greater than that of the binding energy of material exterior to periastron by a factor of two in the prograde case, and up to a factor of five in the case of the retrograde encounter. The destructive nature of the encounters indicates that a non-linear treatment is essential in all but the most distant encounters.

## 1 INTRODUCTION

Much of the recent work on binary star formation has stressed the importance of interactions between stars and their associated gaseous protostellar discs. A proposed mechanism for the formation of binary and multiple systems is that of 'prompt initial fragmentation' (Pringle 1989) in which stars form in bound groups, each comprising a handful of protostellar objects at separations of the order a Jeans length. Each fragment then undergoes the processes of single star formation to yield a central star surrounded by a protostellar disc. At the same time the stars interact under their mutual gravitation. The weak clustering seen after a few million years in star-forming regions is consistent with such a formation mechanism. Optical observations of Taurus–Auriga (Gomez et al. 1993) reveal six such clusters containing approximately 15 stellar systems in each. Speckle imaging surveys of the same region (Ghez et al. 1993) suggest that a large fraction ( $\gtrsim 60$ per cent) of T Tauri stars have companions at *projected* separations comparable with typical disc radii, implying that star–disc interactions may have been important in their formation. In this paper we present a new method for calculating the energy and angular momentum transfer in such star–disc encounters and discuss the implications of our results for binary star formation.

It should be noted that collision rates determined from number densities and velocity dispersions observed in present-day clusters are rather low, and we must therefore appeal to an earlier epoch in the cluster's formation history when number densities were higher, velocity dispersions were smaller, and collision rates were substantial. Based on the prompt formation scenario outlined above, one would expect stellar number densities to be around a few$\times 10^3$ pc$^{-3}$ and velocity dispersions in the sub-1 km s$^{-1}$ regime. Also, the mass of a protostellar disc is much larger during the earliest stages of evolution since much of the enshrouding material has yet to fall in, and thus it is at these earliest stages that discs have the greatest effect on the orbital evolution of the system. The presence of a massive disc offers an enhanced collisional cross-section to nearby members and hence a mechanism for the transfer of orbital energy and angular momentum between perturber orbit and disc material. Penetrating encounters also offer an explanation for the enhanced weak-line T-Tauri binary distribution at small separations (Ghez et al. 1993) by removing much of the material at large radii.

Recent estimates of the rate of binary star formation through star–disc interactions have modelled such encounters in a prescriptive fashion (Clarke & Pringle 1991a, 1991b; Murray, Clarke & Pringle 1991; Larson 1990). For exam-ple, Clarke & Pringle (1991a) took the energy transferred to be equal to the binding energy of the disc material exterior to periastron, and showed, in large-$N$ systems, that even the densest protostellar regions could only yield a binary fraction of up to 20 per cent by this method. In subsequent papers (Clarke & Pringle 1991b; McDonald & Clarke 1995) the authors followed the evolution of a small number of protostars and their discs, experimenting with a number of prescriptions similar to that described above. They found that in small-$N$ clusters the presence of discs assists purely dynamical capture and can give rise to a substantial binary fraction. The binary statistics were, however, somewhat sensitive to the prescription employed, thus underlining the necessity of obtaining a reliable prescription for this process.

In the case of distant encounters, analytical results for energy and angular momentum transfer can be used, these being based on linear perturbation theory. All such theories rely on the perturbing mass to excite waves and resonances in the disc material and calculate the energy and angular momentum transfer induced as a result. Such analysis (Ostriker 1994; Papaloizou & Terquem, in preparation) is similar in both method and results to that on tidal interactions of galaxies. Ostriker calculates energy and angular momentum transfer between a protostellar disc and a perturbing star on a parabolic orbit over a range of disc radius to periastron separation ratios and disc–orbit orientations. She finds that for close encounters disc angular momentum is reduced slightly, and that capture will only occur for polar and retrograde encounters. Papaloizou & Terquem use a similar analysis in computing the angular momentum flux between a disc and a perturbing body and find that the disc loses angular momentum to the perturber and thus accretion on to the central star is increased, and that for non-coplanar encounters the evolution of the system is not necessarily towards coplanarity.

For close and penetrating encounters linear theory is no longer appropriate and different calculational techniques are required. Limited numerical simulations of penetrating star–disc encounters (Clarke & Pringle 1993, hereafter CP93) have shown that outside periastron most of the disc material becomes unbound and thus a suitable estimate of the energy transferred might be the binding energy of the disc material outside periastron. Some attempt was made to account for viscosity using a 'sticky-particle' formulation and it was found that the principal effect of the viscosity was that it enabled re-circularization of the remaining disc material after the fly-by.

Similar studies were conducted by Heller (1993) using Smoothed Particle Hydrodynamics (SPH), where the effect on the stellar orbit was also calculated. To date, this represents the only such calculation in the non-linear regime. This study did not, however, explore the question of which area of the disc is chiefly responsible for transfer of energy and angular momentum to the orbit, and therefore raises the question of whether the critical regions of the disc were adequately resolved. A further uncertainty concerns the inexactitude of the hydrodynamic forces in SPH on a particle by particle basis, a problem that is negligible in many applications where the bulk properties of the flow are of interest, but which may be important in the present problem, if such effects shift particles in and out of critical regions of the disc. On the other hand, a Lagrangian method, such as SPH, is essential in such calculations, as grid-based methods soon founder, during penetrating disc encounters, due to the enormous dynamic range in densities that develop in tidally disturbed regions (Korycansky & Papaloizou 1995).

Faced with these difficulties, and given also the current interest in obtaining reliable star–disc interaction prescriptions, we have opted for an approach that begins by dissecting the star–disc interaction process *in the absence of hydrodynamic forces*. The main motivation of this work is to discover the critical regions of the disc that couple to the stellar orbit and to classify the particle orbits that dominate the energy and angular momentum transfer. Such a study can be used as groundwork for future hydrodynamical calculations, by indicating the regions of the disc that need to be most closely resolved. Furthermore, there are reasons for believing that the results obtained in this ballistic case may not be drastically different from those that would apply when hydrodynamical effects are included: the simulations of CP93 show that the majority of particles in the outer disc are stripped into diverging orbits by the encounter, so that ballistic modelling of the flow is not likely to be greatly erroneous in this case. We therefore, in this paper, examine the ballistic case, since it is clearly analysable on a particle by particle basis and because it permits the economical method (described in Section 2). We bear in mind, however, that our present conclusions may be somewhat modified by gas dynamical calculations.

Our treatment of star–disc interactions in this paper recalls early investigations of interacting galaxies (Toomre & Toomre 1972; Wright 1972; Eneev, Kozlov & Sunyaev 1973; Yabushita 1977) although, due to the computing time required, these studies were mainly qualitative investigations emphasizing the morphology of distorted galaxies. In the present paper we use a new method for calculating the energy and angular momentum transferred in a star–disc encounter, based on the reduced three-body problem. Since we are ignoring all inter-particle interactions in this model, as well as the effect of the disc material on the binary orbit during the numerical experiment, particles may be integrated one at a time. At the end of the experiment we calculate the energy and angular momentum changes of all the particles and equate this to the corresponding change in the binary orbit. Section 2 lays out the basis for this approach, whilst Section 3 outlines the numerical method and Section 4 describes our results. In Section 5 we apply our results to some typical systems, and we draw our conclusions in Section 6.

## 2 DERIVATION OF REDUCED THREE-BODY RESULTS

Consider a system of three masses $M_2$, $M_1$ and $m$ (a typical disc particle) located at $\mathbf{R}_2$, $\mathbf{R}_1$ and $\mathbf{R}$ and moving with velocities $\mathbf{V}_2$, $\mathbf{V}_1$ and $\mathbf{V}$ with respect to the centre of mass of the system. We may then write the total angular momentum, $\mathbf{J}$, of the system as

$$\mathbf{J} = M_2(\mathbf{R}_2 \times \mathbf{V}_2) + M_1(\mathbf{R}_1 \times \mathbf{V}_1) + m(\mathbf{R} \times \mathbf{V}) \qquad (1)$$

and the total energy, $E$, as

$$E \quad = \quad \frac{1}{2}M_2\mathbf{V}_2^2 + \frac{1}{2}M_1\mathbf{V}_1^2 - \frac{GM_2M_1}{\mid \mathbf{R}_2 - \mathbf{R}_1 \mid} + \frac{1}{2}m\mathbf{V}^2$$

$$- \frac{GM_1 m}{\mid \mathbf{R}_1 - \mathbf{R} \mid} - \frac{GM_2 m}{\mid \mathbf{R}_2 - \mathbf{R} \mid}. \qquad (2)$$

We need to know the changes in the energy and angular momentum of the binary orbit as it evolves under the influence of the surrounding disc material. Since we ignore the effect of the disc material on the binary orbit, we must evaluate the changes in the energy and angular momentum of the disc material itself. These may then be equated, with the appropriate change of sign, to the change in energy and angular momentum of the binary orbit, since *total* energy and angular momentum must be conserved. If this is calculated directly, however, we run into problems. Consider the case when a disc particle, $m$, is bound to one of the stars, say $M_1$, while the other star, $M_2$, recedes. The angular momentum of the particle in the centre of mass frame is given by $m(\mathbf{R}\times\mathbf{V})$. As $|\mathbf{R}|\to\infty$, $\mathbf{V}$ tends to the orbital velocity of $m$ about $M_1$ which is finite, and, for a general non-circular orbit, periodically variable in both magnitude and direction. Thus, as $M_2$ recedes, the disc particle's angular momentum oscillates with a frequency equal to that of the $m-M_1$ system and grows in amplitude. Since the disc particle's angular momentum is clearly time-dependent, we cannot guarantee a correct value for this (or other similar quantities) at the end of the simulation. A method which removes this inconvenience will now be described.

Initially, all disc particles are substantially closer to $M_1$ than to $M_2$ (assuming a suitable starting time has been chosen) whilst, after the fly-by, each disc particle may be in one of three situations.

(i) Mass $m$ remains substantially closer to $M_1$ than to $M_2$.
(ii) Mass $m$ is substantially closer to $M_2$ than $M_1$.
(iii) Mass $m$ is substantially further from $M_1$ and $M_2$ than they are from each other.

In each case we can split the three-body system into two 'two-body' subsystems – the close pair of masses becomes one system and its centre of mass and the third mass become the other. By writing the energy and angular momentum in a way appropriate to each of the three possible situations we are able to obtain correct values for the energy and angular momentum of the system.

## 2.1 Case (i): mass $m$ remains substantially closer to $M_1$ than to $M_2$

We first define some convenient coordinates. Let $\mathbf{r}_c$ and $\mathbf{v}_c$ be the position and velocity of the centre of mass of the $m-M_1$ system defined by

$$(M_1+m)\mathbf{r}_c = M_1\mathbf{R}_1 + m\mathbf{R}, \qquad (3a)$$

$$(M_1+m)\mathbf{v}_c = M_1\mathbf{V}_1 + m\mathbf{V}, \qquad (3b)$$

and $\mathbf{r}_1, \mathbf{r}$ and $\mathbf{v}_1, \mathbf{v}$ be the positions and velocities of $M_1$ and $m$ in the centre of mass of the $m-M_1$ system; thus

$$\mathbf{R}_1 = \mathbf{r}_c + \mathbf{r}_1, \qquad (4a)$$

$$\mathbf{R} = \mathbf{r}_c + \mathbf{r}, \qquad (4b)$$

$$\mathbf{V}_1 = \mathbf{v}_c + \mathbf{v}_1, \qquad (4c)$$

$$\mathbf{V} = \mathbf{v}_c + \mathbf{v}. \qquad (4d)$$

Combining (3) and (4) gives the obvious

$$M_1\mathbf{r}_1 + m\mathbf{r} = \mathbf{0}, \qquad (5a)$$

$$M_1\mathbf{v}_1 + m\mathbf{v} = \mathbf{0}. \qquad (5b)$$

Equation (1) now becomes

$$\begin{aligned}\mathbf{J}^{(i)} &= M_2(\mathbf{R}_2\times\mathbf{V}_2) + M_1\left[(\mathbf{r}_c+\mathbf{r}_1)\times(\mathbf{v}_c+\mathbf{v}_1)\right] \\ &+ m\left[(\mathbf{r}_c+\mathbf{r})\times(\mathbf{v}_c+\mathbf{v})\right]. \end{aligned}\qquad (6)$$

Rearranging and noting (5) yield

$$\mathbf{J}^{(i)} = \mathbf{J}_0^{(i)} + \mathbf{J}_1^{(i)} \qquad (7)$$

where

$$\mathbf{J}_0^{(i)} = M_2(\mathbf{R}_2\times\mathbf{V}_2) + (M_1+m)(\mathbf{r}_c\times\mathbf{v}_c), \qquad (8a)$$

$$\mathbf{J}_1^{(i)} = M_1(\mathbf{r}_1\times\mathbf{v}_1) + m(\mathbf{r}\times\mathbf{v}). \qquad (8b)$$

$\mathbf{J}_0^{(i)}$ is the angular momentum of $M_2$ and of the centre of mass of the $M_1-m$ system treated as a point mass in the centre of mass frame.

$\mathbf{J}_1^{(i)}$ is the angular momentum of $M_1$ and $m$ about their mutual centre of mass. Note that these two terms are proportional to $M_1 m^2$ and $M_1^2 m$ respectively and thus in the limit that $m \ll M_1$ only the latter need be considered.

Equation (2) now becomes

$$\begin{aligned}E^{(i)} &= \frac{1}{2}M_2\mathbf{V}_2^2 + \frac{1}{2}M_1\left[\mathbf{v}_c - \frac{m}{M_1}\mathbf{v}\right]^2 - \frac{GM_1M_2}{\mid \mathbf{r}_c - \mathbf{R}_2 + \mathbf{r}_1 \mid} \\ &+ \frac{1}{2}m\left[\mathbf{v}_c - \frac{M_1}{m}\mathbf{v}_1\right]^2 - \frac{GM_1 m}{\mid \mathbf{r}_1 - \mathbf{r} \mid} \\ &- \frac{GM_2 m}{\mid \mathbf{r}_c - \mathbf{R}_2 + \mathbf{r} \mid}. \end{aligned}\qquad (9)$$

Rearrangement yields

$$\begin{aligned}E^{(i)} &= \frac{1}{2}M_2\mathbf{V}_2^2 + \frac{1}{2}(M_1+m)\mathbf{v}_c^2 + \frac{1}{2}\frac{m^2\mathbf{v}^2}{M_1} + \frac{1}{2}\frac{M_1^2\mathbf{v}_1^2}{m} \\ &- \frac{GM_1M_2}{\mid \mathbf{r}_c - \mathbf{R}_2 + \mathbf{r}_1 \mid} - \frac{GM_1 m}{\mid \mathbf{r}_1 - \mathbf{r} \mid} \\ &- \frac{GM_2 m}{\mid \mathbf{r}_c - \mathbf{R}_2 + \mathbf{r} \mid}. \end{aligned}\qquad (10)$$

We now use the condition that $M_1$ and $m$ are much closer to each other than either is from $M_2$, or, more strictly, that $|\mathbf{r}|, |\mathbf{r}_1| \ll |\mathbf{r}_c - \mathbf{R}_2|$. Thus we can expand the fifth and seventh terms to give:

$$E^{(i)} = E_0^{(i)} + E_1^{(i)} \qquad (11)$$

where

$$E_0^{(i)} = \frac{1}{2}M_2\mathbf{V}_2^2 + \frac{1}{2}(M_1+m)\mathbf{v}_c^2 - \frac{GM_2(M_1+m)}{\mid \mathbf{r}_c - \mathbf{R}_2 \mid}, \qquad (12a)$$

$$E_1^{(i)} = \frac{1}{2}\frac{m^2\mathbf{v}^2}{M_1} + \frac{1}{2}\frac{M_1^2\mathbf{v}_1^2}{m} - \frac{GM_1 m}{\mid \mathbf{r}_1 - \mathbf{r} \mid}. \qquad (12b)$$

$E_0^{(i)}$ is the total energy of $M_2$ and of the $M_1-m$ system treated as a point mass. $E_1^{(i)}$ is the 'internal' energy of the $M_1-m$ system. The first two terms in (12b) are proportional to $m^2M_1$ and $mM_1^2$ respectively and thus in the limit that $m \ll M_1$ only the latter need be considered.

### 2.2 Case (ii): mass $m$ is substantially closer to $M_2$ than to $M_1$

The second case is identical to the first but for an interchange of subscripts, thus:

$$\mathbf{J}^{(\mathrm{ii})} = \mathbf{J}_0^{(\mathrm{ii})} + \mathbf{J}_1^{(\mathrm{ii})}, \qquad (13)$$

where

$$\mathbf{J}_0^{(\mathrm{ii})} = M_1(\mathbf{R}_1 \times \mathbf{V}_1) + (M_2 + m)(\mathbf{r}_c \times \mathbf{v}_c), \qquad (14a)$$

$$\mathbf{J}_1^{(\mathrm{ii})} = M_2(\mathbf{r}_2 \times \mathbf{v}_2) + m(\mathbf{r} \times \mathbf{v}), \qquad (14b)$$

and

$$E^{(\mathrm{ii})} = E_0^{(\mathrm{ii})} + E_1^{(\mathrm{ii})}; \qquad (15)$$

here

$$E_0^{(\mathrm{ii})} = \frac{1}{2}M_1\mathbf{V}_1^2 + \frac{1}{2}(M_2+m)\mathbf{v}_c^2 - \frac{GM_1(M_2+m)}{\mid \mathbf{r}_c - \mathbf{R}_1 \mid}, \qquad (16a)$$

$$E_1^{(\mathrm{ii})} = \frac{1}{2}\frac{m^2\mathbf{v}^2}{M_2} + \frac{1}{2}\frac{M_2^2\mathbf{v}_2^2}{m} - \frac{GM_2 m}{\mid \mathbf{r}_2 - \mathbf{r} \mid}. \qquad (16b)$$

Note that the definitions of $\mathbf{v}_c, \mathbf{r}_c, \mathbf{v}\, and\, \mathbf{r}$ change accordingly and that $\mathbf{v}_2$ and $\mathbf{r}_2$ have the obvious meaning. Also, the appropriate terms may be omitted provided that $m \ll M_2$.

### 2.3 Case (iii): mass $m$ is substantially further from $M_1$ and $M_2$ than they are from each other

Case (iii) is a little different.

Define $\mathbf{r}_c$ and $\mathbf{v}_c$ to be the position and velocity of the centre of mass of the $M_1 - M_2$ system defined by

$$(M_1 + M_2)\mathbf{r}_c = M_1\mathbf{R}_1 + M_2\mathbf{R}_2, \qquad (17a)$$

$$(M_1 + M_2)\mathbf{v}_c = M_1\mathbf{V}_1 + M_2\mathbf{V}_2, \qquad (17b)$$

and $\mathbf{r}_1, \mathbf{r}_2$ and $\mathbf{v}_1, \mathbf{v}_2$ to be the positions and velocities of $M_1$ and $M_2$ in the centre of mass of the $M_1 - M_2$ system; thus

$$\begin{aligned} \mathbf{R}_1 &= \mathbf{r}_c + \mathbf{r}_1, & (18a)\\ \mathbf{R}_2 &= \mathbf{r}_c + \mathbf{r}_2, & (18b)\\ \mathbf{V}_1 &= \mathbf{v}_c + \mathbf{v}_1, & (18c)\\ \mathbf{V}_2 &= \mathbf{v}_c + \mathbf{v}_2. & (18d) \end{aligned}$$

Combining (17) and (18) gives the obvious

$$M_1\mathbf{r}_1 + M_2\mathbf{r}_2 = \mathbf{0}, \qquad (19a)$$

$$M_1\mathbf{v}_1 + M_2\mathbf{v}_2 = \mathbf{0}. \qquad (19b)$$

Equation (1) now becomes

$$\begin{aligned} \mathbf{J}^{(\mathrm{iii})} &= M_2\left[(\mathbf{r}_c + \mathbf{r}_2) \times (\mathbf{v}_c + \mathbf{v}_2)\right] \\ &+ M_1\left[(\mathbf{r}_c + \mathbf{r}_1) \times (\mathbf{v}_c + \mathbf{v}_1)\right] + m(\mathbf{R} \times \mathbf{V}). \qquad (20) \end{aligned}$$

Rearranging and noting (19) yield

$$\mathbf{J}^{(\mathrm{iii})} = \mathbf{J}_0^{(\mathrm{iii})} + \mathbf{J}_1^{(\mathrm{iii})} \qquad (21)$$

where

$$\mathbf{J}_0^{(\mathrm{iii})} = M_2(\mathbf{r}_2 \times \mathbf{v}_2) + M_1(\mathbf{r}_1 \times, \mathbf{v}_1), \qquad (22a)$$

$$\mathbf{J}_1^{(\mathrm{iii})} = (M_1 + M_2)(\mathbf{r}_c \times \mathbf{v}_c) + m(\mathbf{R} \times .\mathbf{V}). \qquad (22b)$$

$\mathbf{J}_0^{(\mathrm{iii})}$ is the total angular momentum of the $M_1 - M_2$ system in its centre of mass frame. $\mathbf{J}_1^{(\mathrm{iii})}$ is the angular momentum of $m$ and the $M_1 - M_2$ system treated as a point mass in the centre of mass frame. The two terms in (22b) are proportional to $Mm^2$ and $M^2m$ respectively (where $M = M_1 + M_2$) and in the limit that $m \ll M$ only the latter need be considered.

Equation (2) now becomes

$$\begin{aligned} E^{(\mathrm{iii})} &= \frac{1}{2}M_2(\mathbf{v}_c + \mathbf{v}_2)^2 + \frac{1}{2}M_1(\mathbf{v}_c + \mathbf{v}_1)^2 - \frac{GM_1M_2}{\mid \mathbf{r}_1 - \mathbf{r}_2 \mid} \\ &+ \frac{1}{2}m\mathbf{V}^2 - \frac{GM_1 m}{\mid \mathbf{r}_c + \mathbf{r}_1 - \mathbf{R} \mid} - \frac{GM_2 m}{\mid \mathbf{r}_c + \mathbf{r}_2 - \mathbf{R} \mid}. \qquad (23) \end{aligned}$$

Rearrangement yields

$$\begin{aligned} E^{(\mathrm{iii})} &= \frac{1}{2}(M_1 + M_2)\mathbf{v}_c^2 + \frac{1}{2}M_2\mathbf{v}_2^2 + \frac{1}{2}M_1\mathbf{v}_1^2 \\ &- \frac{GM_1M_2}{\mid \mathbf{r}_1 - \mathbf{r}_2 \mid} + \frac{1}{2}m\mathbf{V}^2 - \frac{GM_1 m}{\mid \mathbf{r}_c + \mathbf{r}_1 - \mathbf{R} \mid} \\ &- \frac{GM_2 m}{\mid \mathbf{r}_c + \mathbf{r}_2 - \mathbf{R} \mid}. \qquad (24) \end{aligned}$$

We now use the condition that $M_1$ and $M_2$ are much closer to each other than either is to $m$, or, more strictly, that $|\mathbf{r}_1|, |\mathbf{r}_2| \ll |\mathbf{r}_c - \mathbf{R}|$. Thus we can expand the sixth and seventh terms to give:

$$E^{(\mathrm{iii})} = E_0^{(\mathrm{iii})} + E_1^{(\mathrm{iii})} \qquad (25)$$

where

$$E_0^{(\mathrm{iii})} = \frac{1}{2}M_1\mathbf{v}_1^2 + \frac{1}{2}M_2\mathbf{v}_2^2 - \frac{GM_1M_2}{\mid \mathbf{r}_1 - \mathbf{r}_2 \mid}, \qquad (26a)$$

$$E_1^{(\mathrm{iii})} = \frac{1}{2}(M_1 + M_2)\mathbf{v}_c^2 + \frac{1}{2}m\mathbf{V}^2 - \frac{G(M_1 + M_2)m}{\mid \mathbf{r}_c - \mathbf{R} \mid}. \qquad (26b)$$

$E_0^{(\mathrm{iii})}$ is the total energy of the binary orbit in its centre of mass frame. $E_1^{(\mathrm{iii})}$ is the total energy of $m$ and of the $M_1 - M_2$ system treated as a point mass. The first two terms in (26b) are proportional to $Mm^2$ and $mM^2$ respectively and in the limit that $m \ll M$ only the latter need be considered.

In the limit $m \ll M_1$ we may take $E_0^{(\mathrm{i})} \simeq E_0^{(\mathrm{ii})} \simeq E_0^{(\mathrm{iii})}$ and $\mathbf{J}_0^{(\mathrm{i})} \simeq \mathbf{J}_0^{(\mathrm{ii})} \simeq \mathbf{J}_0^{(\mathrm{iii})}$, so therefore the changes in energy and angular momentum of the system are simply equal to the difference between the final and initial values of $E_1$ and $\mathbf{J}_1$ respectively, independent of which approach is required to calculate the initial and final values. These changes may then be directly equated to the changes in the energy and angular momentum of the binary orbit, from which the final binary orbital parameters may be obtained from the standard

$$E = -\frac{GM\mu}{2a}, \qquad (27a)$$

$$J = \mu\sqrt{GMa(1 - e^2)}, \qquad (27b)$$

where $M$ is the total mass, $M = M_1 + M_2$, $\mu$ is the reduced mass of the binary system, $\mu = \frac{M_1M_2}{M}$, $a$ is the semi-major axis and $e$ is the eccentricity.

## 3 NUMERICAL METHOD

The disc particles initially describe Keplerian orbits about the primary, $M_1$. Since the particles are non-interacting, the results are independent of the number of disc particles used, as long as the mass attributed to any one particle is much less than either of the stellar masses. In this limit, the energy and angular momentum changes of a disc particle are proportional to its mass, thus allowing any disc density profile to be fitted later, provided that the aforementioned inequality holds. We must also be sure that there are enough particles throughout the disc to follow any localized regions of interest. Disc particles are massless and thus the response of the binary to the disc is not calculated during the experiment. No form of dissipative process is employed and therefore the disc particles are integrated ballistically in the time-dependent binary potential. The binary consists of two point masses, the primary, $M_1$, and the perturber, $M_2$, on parabolic orbits about their centre of mass. We begin the calculation at a time of $t = t_0$ (t = 0 at periastron) when the binary separation is at least ten times the initial disc radius, $r_{\rm disc}$. The initial positions are found by solving Kepler's equations (Taff 1985). We find that the disc evolution is insensitive to the initial binary separation as long as the initial separation is substantially larger than the initial disc radius.

The motion of both the disc particles and the point masses is integrated using the Bulirsch–Stoer method. We use an unsoftened gravitational potential for each star. The accuracy of this method was checked by integrating the system over the time of a typical fly-by, but using a zero-mass perturber. We found energy conserved to better than one part in $10^9$ and angular momentum to one part in $10^{12}$.

At the end of each timestep we determine into which of the three categories described above each of the particles falls, and $E_1$ and $\mathbf{J}_1$ are calculated appropriately (equations 8b, 12b, 14b, 16b, 22b, and 26b). Only when $E_1$ and $\mathbf{J}_1$ have stabilized and we can be sure that our values for the energy and angular momentum transfer are sufficiently accurate do we stop integration of that particle. Particles are integrated in a serial fashion.

Particles that lie within some chosen radius of $M_1$ or $M_2$ and are gravitationally bound to that mass are considered accreted to that mass and are no longer integrated. Since our chosen accretion radius of 0.5 per cent times the initial disc radius, i.e. $r_{\rm accrete} = 5\times10^{-3} r_{\rm disc}$, is always much less than our interstellar separation we can be sure that the values calculated for $E_1$ and $\mathbf{J}_1$ at the point where we decide the particle has been accreted will be accurate. Thus, even though particles may be accreted early in the simulation, we can confidently include their contribution to the transfer.

## 4 RESULTS

We present the results of our tests of the numerical method and of several models in which we vary the disc radius with respect to $r_{\rm peri}$ and the relative inclination of the disc and orbital planes. In all models $M_2/M_1 = 1$ and the binary orbit is parabolic. Our code time units are

$$\begin{aligned} \tau &= \frac{1}{2}(r_{\rm peri}^3/GM)^{1/2} \\ &= 0.08(r_{\rm peri}/{\rm au})^{3/2}(M/{\rm M}_\odot)^{-1/2}\ {\rm yr}, \end{aligned} \quad (28)$$

where $M = M_1 + M_2$.

Figure 1 shows the axes we use. The primary lies at the origin and the perturber orbits in the $x - y$ plane. The binary angular momentum vector is therefore in the positive $z$ direction. The angular momentum vector of the disc, $\mathbf{J}_{\rm disc}$, always lies in the $y - z$ plane and makes an angle, $\theta$, with that of the binary, $\mathbf{J}_{\rm orbit}$, in the sense indicated. $\theta = 0°$ in the prograde encounter and 180° in the retrograde scenario.

**Figure 1.** We employ axes in which the primary lies at the origin and the perturber orbits in the $x - y$ plane. The binary angular momentum vector, $J_{\rm orbit}$, is therefore in the positive $z$ direction. The disc's angular momentum vector, $J_{\rm disc}$, lies in the $y - z$ plane and makes an angle, $\theta$, with that of the orbit in the sense shown.

### 4.1 The numerical method

The accuracy of the Bulirsch–Stoer method has already been established and we do not go into the details here. In order to ensure that we have enough particles in our disc, we choose a representative ring undergoing a strong interaction with the perturber and monitor the change in its energy and angular momentum as we vary $N$, the number of particles that it contains. We use a ring of radius $1.9 r_{\rm peri}$ and several values for $N$, ranging from 10 to 1000. At low values of $N$, accuracy is very poor, especially in the angular momentum, whilst for $N >$ few$\times10^2$ both energy and angular momentum are within a few per cent of the values obtained with 1000 particles in the ring, an indication that we are populating the disc densely enough to be sure that we are following all of the interaction. Unfortunately, the computational time required to evaluate 1000 particles in this ring implies prohibitively long computational times for an entire disc, and thus we choose $N = 100$ as a compromise between computational speed and numerical accuracy. Having said that, however, when $N \sim 100$, the tolerance in the energy is less than 1 per cent and less than 10 per cent in angular momentum. We have also examined rings at other radii and found that, for similar accuracy, 50 particles are sufficient at $r/r_{\rm peri} = 1.0$ and a mere 13 in the innermost region at $r/r_{\rm peri} = 0.25$.

### 4.2 Coplanar, prograde encounter

#### *4.2.1 Particle distributions*

In the same vein as CP93 we first discuss the results for a coplanar, prograde encounter. The stars are on parabolic orbits and $M_2/M_1 = 1$. Our post-encounter particle distributions are similar to those of CP93 and the fractions of particles that become unbound or are captured (Fig. 2) are also in good agreement with CP93. The section of Fig. 2 for $r/r_{\rm peri} < 0.8$ should be compared to fig. 3 of CP93.

**Figure 2.** The fate of particles as a function of initial radius for a coplanar, parabolic, prograde orbit with the mass ratio $M_2/M_1 = 1$. At each radius within the disc, the fraction of particles at that radius that remain bound to the primary (asterisks), that are captured by the perturber (filled squares) and that become unbound from the system (open squares) are shown.

**Figure 3.** The fraction of disc particles in each of the three final states, bound to the primary (asterisks), captured (filled squares) and unbound (open squares), plotted against time, $t$; $t = 0$ at periastron. The encounter is coplanar, parabolic and prograde with $M_2/M_1 = 1$.

Outside periastron the number of particles that are captured falls to only a few per cent compared with 40 per cent at $r/r_{\rm peri} = 0.6$. The amount of material remaining bound to the primary reaches a minimum of 20 per cent at $r/r_{\rm peri} = 1.3$ which rises again to 40 per cent at the disc's edge. The remainder of the material becomes unbound, the exact amount varying between 60 and 75 per cent, peaking at $r/r_{\rm peri} = 1.5$. The fact that this fraction is so high adds weight to the validity of the assumption made in previous papers (McDonald & Clarke 1995; Clarke & Pringle 1991b) that the reaction is dominated by the unbinding of all the disc material exterior to periastron.

Fig. 3 shows the time evolution of these fractions. The most important thing to notice is that the bulk of the interaction takes place within 1 code unit of periastron ($t = 0$), when the perturber is well within $r_{\rm disc}$ of the primary, and by a time of $t = 5.0$ the final state of the particles is evident. We find that 58 per cent of the disc particles become unbound, 30 per cent remain bound to the primary and the remaining 12 per cent are captured by the perturber. These percentages reflect the number of test particles in our disc (using a uniform surface number density) and *not* disc mass which depends on the disc's surface density profile through the radial dependence of the mass ascribed to each particle.

### *4.2.2 Energy and angular momentum transfer*

Since we find that the majority of particles have settled into their final bound or unbound states very soon after periastron passage, further integration is only necessary to ensure that we have accurate values for the energy and angular momentum according to Section 2. We find that, by a time $t = 300$, over 90 per cent of particles have been integrated for a sufficient length of time to ensure that their energies and angular momenta have settled to within the tolerance level specified (one part in $10^3$). The remaining 10 per cent take a little longer to complete, although not prohibitively so. Fig. 4 shows the particles in their initial positions with respect to the primary at {0,0}. The magnitudes of the energy and angular momentum changes are represented by the area of the symbol.

We find, as we might expect, some structure in the pattern of energy transfer over the disc's area. In the central regions ($r/r_{\rm peri} \lesssim 0.3$) there is very little change, whilst at intermediate radii ($0.3 \lesssim r/r_{\rm peri} \lesssim 1.5$) the energy transfer is fairly uniformly distributed. At large radii ($r/r_{\rm peri} \gtrsim 1.5$) we find there is a significant excess confined to a small range in azimuthal angle, this range being different at different radii. It is clear that most of the interaction is not taking place in one small region of the disc and thus there is no need to overpopulate parts of the disc selectively.

The peaks in the energy transfer as shown in Fig. 4(a), form a spiral pattern and one might expect these peaks to become aligned at some point due to the effects of differential rotation associated with the Keplerian nature of the disc. However, the picture is somewhat more complicated than that. Selecting the dominant particles only (Figure 5a), and following their motions as they approach periastron, we found that they form four distinct groups (Fig. 5b). Group 1 is composed of approximately 20 per cent of these particles, which undergo close encounters with the perturber as their

**Figure 4.** The locations of disc particles in the disc at the start of the simulation. The area of each symbol is proportional to the absolute magnitude of (a – top) the energy and (b – bottom) the angular momentum changes that the particle suffers during the encounter. The encounter is coplanar, parabolic and prograde with $M_2/M_1 = 1$.

**Figure 5.** (a – top) The peaks in energy transfer, selected from Fig. 4(a). The system is shown at time $t = -40$. (b – bottom) The same system at a time $t = -2$, showing how the particles have evolved into four distinct groups. Groups are labelled left to right – group 1 (open triangles), group 2 (filled triangles), group 3 (open squares) and group 4 (filled squares). The perturber's orbit (indicated by the solid line) is prograde (anti-clockwise here), parabolic and coplanar. The central star and the perturber are represented by the bold, open squares. The area of the symbol is proportional to the magnitude of that particle's energy change.

orbital motion about the primary brings them close to the perturber (Fig. 6a). Group 2 particles, again comprising approximately 20 per cent, undergo close encounters with the perturber as its motion brings it close to them (Fig. 6b). The third is by far the dominant group, comprising 50 per cent of particles. As the perturber approaches the disc, these particles are disrupted from their original orbits and go into orbit about the perturber *in the opposite sense to that of the*

**Figure 6.** The time evolution of the peaks in energy transfer shown in Fig. 5. (a – top left) At time $t = -1.6$. (b – top right) At time $t = -1.1$. (c – bottom left) At time $t = -0.5$. (d – bottom right) At time $t = -0.3$.

*disc about the primary.* They then undergo close encounters with the primary and become unbound (Fig. 6c). The remaining particles form the fourth group and, like those in the third group, are sufficiently perturbed that they begin to orbit the primary in the opposite sense and suffer close encounters with the perturber as it approaches periastron (Fig. 6d). Note that in Figs 6(c) and (d) the particles from groups 1 and 2 can be seen scattered throughout the system and their motion away from the centre of mass of the system is evident. These peaks carry approximately one third of the total energy transferred, suggesting that close encounters of the kind demonstrated here are significant and warrant closer examination. It is clear that this is a highly non-linear process and thus must be treated numerically.

There is no obvious correlation between peaks in energy transfer and peaks in angular momentum transfer throughout the disc, except in the almost undisturbed, central ($r/r_{\rm peri} \lesssim 0.3$) region. It is therefore imperative that we follow all of the disc throughout the interaction, rather than just concentrate on the energy or angular momentum 'hotspots'. Binning particles according to the magnitude of the change in their energy and angular momenta (Fig. 7), we see that the majority of the particles have relatively small changes and that these particles produce the dominant effect (Fig. 8). This is reassuring in as much as, as has already been said, there is no one small region of the disc, and thus only a few particles, that dominates the interaction.

We can see clearly that interior to $r/r_{\rm peri} \approx 0.3$ there is almost no perturbation to the disc material. Fig. 9(a) shows the energy change at each radius. Between $r/r_{\rm peri} \approx 0.3$ and $\approx 0.6$ the energy change is negative; the material between these radii is becoming more tightly bound and, correspondingly, the binary is becoming more weakly bound. Thus, if our disc were not to extend farther than $0.6 r_{\rm peri}$ (this value depending on the mass ratio $M_2/M_1$, and the initial binary orbital parameters), we would expect the binary system to gain energy. Note that co-rotation between the perturber and disc material occurs at $r/r_{\rm peri} = 0.63$ at periastron and we suspect that the negative-going part of Fig. 9(a) may be a co-rotation resonance feature. It has been pointed out, however, that an $m = 2$ inner Lindblad resonance (occurring at $r/r_{\rm peri} = 0.4$) is also possible (E.C. Ostriker, private communication). We are unable to distinguish between the two with our current results. It is interesting to note that the inner part of this feature is dominated by material that remains bound to the primary – at this radius approximately 95 per cent of the material remains bound – whilst the outer part of the resonance is dominated by material that is captured by the perturbing mass. At this radius ($r/r_{\rm peri} \approx 0.5$) approximately one third of the material is captured (Fig. 2). Thus even when $r_{\rm peri}/r_{\rm disc} \sim 2$ the interaction is already non-linear and can no longer be treated in a linear fashion.

**Figure 7.** The particles are binned according to the magnitude of their energy and angular momentum changes during a coplanar, prograde, parabolic encounter with $M_2/M_1 = 1$. The numbers of particles in each bin are plotted. The solid line indicates binning by energy, the dashed line by angular momentum.

**Figure 8.** The particles are binned according to the magnitude of their energy and angular momentum changes during a coplanar, prograde, parabolic encounter with $M_2/M_1 = 1$. The total contributions to the energy and angular momentum transfer from the particles in each bin are plotted. The solid line indicates binning by energy, the dashed line by angular momentum.

Beyond $r/r_{\rm peri} \approx 0.6$, the curve rises sharply to a peak just inside periastron. In this region, as one might expect given the destructive nature of the interaction, the reaction is now dominated by the material that becomes unbound from the system. There is some residual effect from material that remains bound to the system, acting in the opposite direction, i.e. it loses energy to the binary orbit. In the region $1 \lesssim r/r_{\rm peri} \lesssim 2$ the total energy transferred is approximately constant, coinciding with a peak in the fraction of material becoming unbound from the system (Fig. 2). Beyond $r/r_{\rm peri} \approx 2$, the total energy drops as approximately $r^{-1}$, being almost wholly due to the unbound material; material that ends up being bound to either star contributes very little to the total energy transfer. The fluctuations present in the curve in this region have been shown to be $\sqrt{N}$ noise. At these radii, the $r^{-1}$ dependence of the energy is reflected in the fraction of material becoming unbound (Fig. 2). These results agree with those of Yabushita (1977) who found that material at radii $r/r_{\rm peri} = 0.375$ and $r/r_{\rm peri} = 0.5$ accelerated the relative motion of the perturber whilst material at $r/r_{\rm peri} = 0.625$ tended to decelerate the relative motion of $M_2$ and $M_1$.

Fig. 9(b) shows the equivalent results for angular momentum. Again within $r/r_{\rm peri} \approx 0.3$ the transfer is negligible. The same resonance feature is present between $r/r_{\rm peri} \approx 0.3$ and $r/r_{\rm peri} \approx 0.6$ where the disc material is losing angular momentum to the binary orbit – the structure of this feature is as for the energy transfer. Outside $r/r_{\rm peri} \approx 0.6$ the angular momentum transfer rises sharply, again dominated by the unbound material. The total transfer is approximately constant out to $r/r_{\rm peri} \approx 1.5$ where it falls equally sharply to zero at $r/r_{\rm peri} \approx 2$. Here, the transfer is equally dominated by the unbound material gaining angular momentum and the bound material losing angular momentum. Beyond $r/r_{\rm peri} \approx 2$, the total angular momentum transfer becomes increasingly negative with increasing $r$, only it is now dominated by the material that remains bound. This is coincidental with an increasing fraction of disc material remaining bound at these radii (Fig. 2) and reinforces our earlier findings showing no correlation between peaks in energy and angular momentum transfer. The angular momentum transfer at large radii is dominated by material that suffers a reduction in its angular momentum whilst remaining bound. Its contribution to the energy transfer at these radii is small, however, due to its small initial energy. There is a 'peak' in the total angular momentum transfer at $r/r_{\rm peri} \approx 2.8$ which is also present in the transfer from the material bound to the primary. There is no evidence that this is a resonance feature. Captured material is also important at large radii, its effect to some extent cancelling that of the unbound material.

**Figure 9.** The energy change (a – top) and angular momentum change (b – bottom) per unit mass plotted against radius for the coplanar, parabolic, prograde encounter with $M_2/M_1 = 1$. The contributions from primary-bound material (asterisks), captured material (filled squares) and unbound material (open squares) are plotted along with the total (open triangles).

It would appear that by $r/r_{\rm peri} \gtrsim 3$ the transfer curves are tending to the limiting values at large radius. Calculations with a ring of material at $r/r_{\rm peri} = 10$ confirm this.

**Figure 10.** The fate of particles as a function of initial radius for a coplanar, parabolic, retrograde orbit with the mass ratio $M_2/M_1 = 1$. At each radius within the disc, the fractions of particles at that radius that remain bound to the primary (asterisks), that are captured by the perturber (filled squares) and that become unbound from the system (open squares) are shown.

The fractions falling into the three final configurations are as follows: unbound 26 per cent, bound to the primary 52 per cent and captured by the secondary 22 per cent. Similarly the energy transfers for each of the three cases are tending towards zero. In the case of the angular momentum, things are not quite so clear. The contributions from primary and secondary bound material remain at values similar to those at $r/r_{\rm peri} = 4$ whilst the transfer due to the unbound material changes sign although the magnitude remains approximately the same. The consequence of this is that the total angular momentum transfer maintains an approximately linear increase in magnitude to a value of $-1.35$ at $r/r_{\rm peri} = 10$.

### 4.3 Coplanar, retrograde encounter

In this encounter we now have the disc rotating in the opposite sense to the perturber about the primary star. In this case we not expect to find a resonance feature, for obvious reasons.

#### *4.3.1 Particle distributions*

Again, our post-encounter particles distributions are in good agreement with those of CP93. Overall (using a constant surface density profile), we find that 62 per cent of the disc particles become unbound from the system during the encounter, 28 per cent remain bound to the primary and the remaining 10 per cent are captured by the perturbing body. Fig. 10 shows how the fractions of particles in each of the three final states vary with the initial radius. Within $r/r_{\rm peri} \approx 0.7$ all disc particles remain bound to the primary. Outside this, the number that become unbound rises sharply, peaking at 75 per cent at $r/r_{\rm peri} \approx 1.6$. Correspondingly, the fraction that remain bound drops sharply, reaching a minimum of 25 per cent at the same radius. This figure should be compared with fig. 8 of CP93 for the region $r/r_{\rm peri} < 2$. Beyond $r/r_{\rm peri} \approx 2$ some material is captured, the fraction increasing linearly with radius, reaching 25 per cent at $r/r_{\rm peri} = 4$. The fraction remaining bound to the primary drops with increasing radius in this region, reaching 20 per cent at $r/r_{\rm peri} = 4$. Correspondingly, the fraction becoming unbound remains approximately constant exterior to $r/r_{\rm peri} \approx 2$ at approximately 65 per cent.

**Figure 11.** The locations of disc particles in the disc at the start of the simulation. The area of each symbol is proportional to the absolute magnitude of (a – top) the energy and (b – bottom) the angular momentum changes that the particle suffers during the encounter. The encounter is coplanar, parabolic and retrograde with $M_2/M_1 = 1$.

**Figure 12.** (a – top) The peaks in energy transfer, selected from Fig. 11(a). The system is shown at time $t = -40$. (b – bottom) The same system at a time $t = -2$, showing how the particles have evolved into three distinct groups. Groups are labelled 1 to 3 in a clockwise fashion starting from the left – group 1 (open triangles), group 2 (filled triangles) and group 3 (open squares). The perturber's orbit (indicated by the solid line) is parabolic and coplanar. Disc material orbits in a retrograde fashion (clockwise here). The central star and the perturber are represented by the bold, open squares. The area of the symbol is proportional to the magnitude of that particle's energy change.

**Figure 13.** The time evolution of the peaks in energy transfer shown in Fig. 12. (a – top left) At time $t = -1.2$. (b – top right) At time $t = 0.5$. (c – bottom) At time $t = 0.8$.

### *4.3.2 Energy and angular momentum transfer*

Fig. 11 shows how the energy and angular momentum transfers are distributed throughout the material. Again, there is a noticeable pattern in the energy transfer: the inner, undisturbed region is larger than before, extending to $r/r_{\rm peri} \approx 0.8$. In the region $0.8 \lesssim r/r_{\rm peri} \lesssim 2$ there is a large amount of energy transfer whilst at larger radii the peaks in energy transfer are localized within small ranges of azimuthal angle, different at different radii. These peaks form a spiral structure as was present in the prograde case, but of opposite sense here. This again suggests an alignment of these peaks via the differential rotation of the disc (clockwise in the retrograde case).

Selecting these peaks (Fig. 12a) we find that they form not one but three distinct clumps by a time $t = -2$ (Fig. 12b). The third group is the most dominant of the three, consisting of approximately 40 per cent of the most dominant particles. Their motion around the central star brings them close to the perturber as it traverses the disc where they undergo close encounters with the perturber, being scattered in the process. It should be noted that, due to the proximity of their encounters, none is scattered through angles less then $\pi/2$ (Fig. 13a). The second group comprises around half of the particles which deviate from their orbits around the primary due to the presence of the perturber and undergo close encounters with the primary (Fig. 13b). The first group is much smaller, comprising only 10 per cent of the particles which are scattered by the perturber as it passes through the disc on its way away from the central star – again there are no scatterings through an angle less than $\pi/2$.

As for the prograde case, we find no correlation between peaks in energy and angular momentum transfer, except for the relatively undisturbed inner region. The angular momentum transfer is spread approximately uniformly over the region exterior to $r/r_{\rm peri} = 2$.

Fig. 14(a) shows the energy transfer as a function of radius for the coplanar, retrograde encounter. The undisturbed region near the primary now extends to $r/r_{\rm peri} \approx 0.8$. We expect this since the relative velocity of the particles with respect to the perturber in this region is high, so the time available for the interaction is short as well as the separation being large. We see no resonance feature, as expected. This undisturbed region confirms the calculations of Yabushita (1977) who found energy changes caused by material interior to $r/r_{\rm peri} = 0.625$ to be insignificant. In the region $0.6 \lesssim r/r_{\rm peri} \lesssim 1$ there is a small blip in the energy transfer performed by the primary-bound material. In this region, however, the energy transfer rises very steeply to a peak at $r/r_{\rm peri} \approx 1.2$; the energy transfer at this maximum being over twice that at the maximum in the coplanar case. The energy transfer is approximately constant over the range $1.2 \lesssim r/r_{\rm peri} \lesssim 1.6$. At large radii the energy transfer falls off as approximately $r^{-2}$. Throughout, the total energy transfer is overwhelmingly dominated by the unbound material. Even though an increasing fraction is being captured at large radii (Fig. 10), its effect is negligible throughout the range examined.

Fig. 14(b) shows the equivalent result for the angular momentum transfer. Note that the initial angular momentum of the binary is positive, that of the disc is negative, and so the angular momentum of the disc material is increasing in the direction of that of the orbit. Again, the inner region ($r/r_{\rm peri} \lesssim 0.8$) is undisturbed. At $r/r_{\rm peri} \approx 0.8$, the angular momentum transfer rises steeply and is dominated by unbound material. The rise becomes less steep at periastron and continues approximately linearly with radius towards large radii. Outside periastron the transfer is still dominated by the unbound material although not as completely as at periastron. Exterior to $r/r_{\rm peri} \approx 1.6$ there is a significant contribution from the material remaining bound to the central star, remaining approximately constant in this region. At large radii ($r/r_{\rm peri} \gtrsim 3$), there is an increasing contribution from captured material, corresponding to the increased fraction of material being captured at these radii.

### 4.4 Angle of encounter

While, for a binary system forming out of a single rotating cloud or from disc fragmentation, one might expect there to be a correlation between the angular momenta of the disc and of the binary orbit, we may not get such a correlation in binaries formed during the collapse of a cluster of stars. Surveys of solar-type binary systems reveal similar correlations (Hale 1994, for example) which show that for binary systems there is a correlation between the equatorial plane and the orbital plane only for $a \lesssim 30$ au and that there is no preferred orientation in hierarchical systems. It is therefore important that non-coplanar encounters are considered.

We now consider three different relative orientations of the disc and orbit, $\theta = 45°, 90°$, and $135°$, where prograde encounter is at $\theta = 0°$ and retrograde is at $\theta = 180°$. In all cases the binary's major axis lies in the disc plane (Fig. 1).

We recognize the importance of encounters where periastron does not lie in the disc plane, but due to space restrictions we reserve these for a follow-up paper. However, following some preliminary calculations, we do not expect a substantial difference between orientations with periastron outside the disc plane and those presented here.

**Figure 14.** The energy change (a – top) and angular momentum change (b – bottom) per unit mass plotted against radius for the coplanar, parabolic, retrograde encounter with $M_2/M_1 = 1$. The contributions from primary-bound material (asterisks), captured material (filled squares) and unbound material (open squares) are plotted along with the total (open triangles).

**Figure 15.** Energy (a – top left) and the three components ($x$ – (b – top right), $y$ – (c – bottom left), $z$ – (d – bottom right)) of the angular momentum transfer per unit mass in the three non-coplanar encounters: 45° open triangles, 90° open squares and 135° open circles. The angular momentum of the original binary orbit is in the positive $z$ direction.

#### *4.4.1 Particle distributions*

Using a constant surface density profile in all three cases we find the fractions of disc particles that end up in the three possible situations are as follows. In the $\theta = 45°$ encounter, we find that, overall, approximately 33 per cent of the disc particles become unbound from the system, 32 per cent remain bound to the primary and 35 per cent are captured by the perturber. For $\theta = 90°$, 57 per cent of the disc particles become unbound from the system during the encounter, 36 per cent remain bound to the primary and 7 per cent are captured. These fractions are somewhat similar to those for the $\theta = 135°$ case at 53, 46 and 1 per cent respectively.

#### *4.4.2 Energy transfer*

In all three encounter orientations we can see that, as before, the inner regions are only slightly disturbed (Fig. 15). For $\theta = 45°$, a resonance feature is present although at a slightly different radius from the prograde case. This may be expected since we must now consider the projection of the motion of the disc material on to the plane of the binary orbit. For $\theta = 45°$ co-rotation occurs at $r/r_{\rm peri} = 0.5$ and the $m = 2$ inner Lindblad resonance at $r/r_{\rm peri} = 0.3$. There is a small negative-going region at $r/r_{\rm peri} = 0.5$ in the $\theta = 90°$ encounter, probably caused by some warping of the inner disc plane during fly-by that enables a small resonance effect to occur.

Exterior to $r/r_{\rm peri} \sim 0.6$, the energy transferred rises sharply to a peak exterior to periastron, in the range $1 \lesssim r/r_{\rm peri} \lesssim 2$, before falling off at larger radii. The magnitude of the peak in energy transfer is similar for all three encounter angles and comparable to that in the prograde case. In all cases the energy transfer is dominated by the material that becomes unbound from the system.

**Figure 16.** The total (a – top) energy change and (b – bottom) angular momentum change of the disc material within a radius $r$ per unit mass within a radius $r$ plotted against $r$ for the coplanar, prograde encounter. The binary orbit is parabolic and $M_2/M_1 = 1$. Each of the three density profiles is shown: constant surface density (open triangles), $\propto r^{-1}$ (open squares), $\propto r^{-7/4}$ (open circles). The filled symbols indicate the binding energy of the material between that radius and $r/r_{\rm peri} = 1$ for the respective profiles.

**Figure 17.** The total (a – top) energy change and (b – bottom) angular momentum change of the disc material within a radius $r$ per unit mass within a radius $r$ plotted against $r$ for the coplanar, retrograde encounter. The binary orbit is parabolic and $M_2/M_1 = 1$. Each of the three density profiles is shown: constant surface density (open triangles), $\propto r^{-1}$ (open squares), $\propto r^{-7/4}$ (open circles). The filled symbols indicate the binding energy of the material between that radius and $r/r_{\rm peri} = 1$ for the respective profiles.

### 4.4.3 Angular momentum transfer

In the $\theta = 45°$ encounter, the angular momentum in the $x$ direction (along which the binary's major axis lies) hardly changes (Fig. 15b). The transfer is dominated in the outer regions ($r/r_{\rm peri} \gtrsim 2.4$) by material whose $y$ component is being reduced (Figure 15c). (The initial disc angular momentum is in the $\hat{\mathbf{y}} + \hat{\mathbf{z}}$ direction.) Interior to $r/r_{\rm peri} \sim 1.5$ the transfer is dominated by material whose $z$ component is increasing (Fig. 15d). In between, both effects are approximately equal.

In the $\theta = 90°$ orientation the disc's initial angular momentum is in the $\hat{\mathbf{y}}$ direction. From Figs 15(b)–(d) it is clear that there is little change in any component in the innermost regions of the disc whilst, at intermediate radii ($0.8 \lesssim r/r_{\rm peri} \lesssim 2$), the transfer is dominated by the $z$ component which is becoming more positive at the expense of the binary's angular momentum (originally in the positive $z$ direction). At large radii ($r/r_{\rm peri} \gtrsim 2$), we find that the $y$ component peaks strongly at $r/r_{\rm peri} \sim 3.3$ (Fig. 15c) and dominates the interaction. The $z$ component undergoes a small change in this region whilst the change in the $x$ component is of intermediate value.

We find that all three components are of approximately equal magnitude at large radii in the $\theta = 135°$ encounter whilst, in the inner regions, changes in the $x$ and $y$ components dominate. The initial disc angular momentum is in the $\hat{\mathbf{y}} - \hat{\mathbf{z}}$ direction.

In all cases material gains angular momentum in the $z$ direction and thus the $z$ component of the initial binary orbital angular momentum is reduced. As was found in the coplanar encounters, the interaction is dominated by material becoming unbound from the system.

## 5 APPLICATION OF RESULTS

The quantitative results we have presented so far have been 'per unit mass', thus allowing any reasonable radial density profile to be assumed later. In order to fit an arbitrary disc density profile, $\Sigma(r)$, it is simply a matter of multiplying the radial energy and angular momentum transfer curves by

$$\Sigma(r)\, 2\pi r\, \Delta r \qquad (29)$$

where $\Delta r$ is the radial extent of each annulus of material. We now apply our results to some 'typical' discs. It should be noted that we still scale our masses by $M_1$ and distances by $r_{\rm peri}$ and any application to real systems will be left as an exercise for the reader. One caveat to this is that in the regime of large disc masses, i.e. $M_{\rm disc}/M_1 \not\ll 1$, this linear relationship no longer applies, not only because the dynamics of the disc would be affected by its self-gravity, but also because the effect of the disc material on the binary orbit would be significant, thus invalidating the reduced three-body approach (Section 2).

We present the results of three density profiles in such a way as to simplify their application to specific systems. A constant surface density was chosen as reference profile and contrasted with the results for two power-law profiles.

Figs 16(a) and (b) show the energy and angular momentum changes for the prograde encounter for each of the four profiles. Plotted vertically is the energy change of the material interior to radius $r$ divided by the mass in the same region. Thus in order to apply a specific disc mass, $M_{\rm disc}/M_1$, and radius, $r/r_{\rm peri}$, one simply reads the energy or angular momentum change per unit mass off the appropriate profile curve at the required value of $r/r_{\rm peri}$ and multiplies by the ratio $M_{\rm disc}/M_1$. Also plotted is the binding energy of the material exterior to periastron as given by

$$E_{\rm bind} = -\int_{r_{\rm peri}}^{r_{\rm disc}} \frac{1}{2}\frac{GM_1}{r}\ \Sigma\ 2\pi r {\rm d}r. \qquad (30)$$

Figs 17(a) and (b) show the equivalent results for the retrograde encounter.

It should be noted that, for the steeper density profiles, the magnitude of the energy change is smaller than for other profiles since a steeper profile results in more of the mass being present in the inner regions where the transfer effects are smaller.

In both prograde and retrograde cases and for all density profiles, the energy transferred is always greater than the binding energy of the disc. The energy transfer is a factor of approximately two larger than the energy transfer in the prograde case, rising to a factor of five in the retrograde case. Thus any assumption that equates the energy loss of a binary orbit to the binding energy of disc material exterior to periastron is *underestimating* the strength of the encounter. This effectively means that the disc masses in previous prescriptive work (Clarke & Pringle 1991a, 1991b; Murray et al. 1991, for example) can be reduced by a factor of between two and five. We stress, however, that the results obtained here for low-mass discs may not necessarily be applicable to the massive discs ($M_{\rm disc} \sim M_1$) considered in these works.

Interior to periastron, the energy and angular momentum changes are far from small and when inside $r/r_{\rm peri} \approx 0.6$, where the resonance feature dominates, the energy and angular momentum of the disc material are actually reduced. It is therefore imperative that the effects of material interior to periastron are accounted for, particularly if periastron lies near the edge of the disc.

Tables 1 and 2 show the results of applying Figs 16 and 17 to some 'typical' disc systems in the prograde and retrograde cases respectively. We choose two typical disc masses, $M_{\rm disc}/M_1 = 10^{-2}$ and $M_{\rm disc}/M_1 = 10^{-1}$, and disc radii, $r_{\rm disc}/r_{\rm peri} = 0.5$ and $r_{\rm disc}/r_{\rm peri} = 2$. In all cases we have calculated the parameters of the final binary orbit as per equations (27). The initial orbital energy and angular momentum take the values $E_{\rm initial} = 0$ (implying a parabolic orbit) and $\mathbf{J}_{\rm initial} = 0.5\hat{\mathbf{z}}$ (indicating periastron separation) respectively.

In all cases, we find that the periastron separation of the final binary orbit is very close to that of the original. The maximum deviation shown in our examples is of 14 per cent for a constant-$\Sigma$ disc with mass $M_{\rm disc}/M_1 = 0.1$ and $r_{\rm disc}/r_{\rm peri} = 2$ in the retrograde case. This, combined with the abrupt nature of the interaction, suggests that an impulse approximation applied at periastron would not be drastically erroneous.

**Table 1.** The application of our results to the three disc density profiles, flat, $\propto r^{-1}$ and $\propto r^{-7/4}$. We use two disc masses and two disc radii. The encounter is coplanar and *prograde* with $M_2/M_1 = 1$. Our initial energy and angular momentum take the values $E_{\rm initial} = 0$ and $J_{\rm initial} = 0.5$ respectively. $a'$ is the semi-major axis, $e'$ is the eccentricity and $r'_{\rm peri}$ is the periastron separation of the post-encounter binary orbit.

| $\Sigma(r)$ | $\frac{r_{\rm disc}}{r_{\rm peri}}$ | $\frac{\Delta E_{\rm orbit}}{M_{\rm disc}}$ | $\frac{\Delta J_{\rm orbit}}{M_{\rm disc}}$ | $\frac{M_{\rm disc}}{M_1} = 10^{-2}$: $\frac{a'}{r_{\rm peri}}$ | $e'$ | $\frac{r'_{\rm peri}}{r_{\rm peri}}$ | $\frac{M_{\rm disc}}{M_1} = 10^{-1}$: $\frac{a'}{r_{\rm peri}}$ | $e'$ | $\frac{r'_{\rm peri}}{r_{\rm peri}}$ |
|---|---|---|---|---|---|---|---|---|---|
| *const.* | 0.5 | $1.43 \times 10^{-1}$ | $6.13 \times 10^{-2}$ | $-3.50 \times 10^{2}$ | 1.0029 | 1.0010 | $-3.50 \times 10^{1}$ | 1.029 | 1.010 |
| | 2 | $-4.57 \times 10^{-1}$ | $-3.22 \times 10^{1}$ | $1.09 \times 10^{2}$ | 0.9909 | 0.9917 | $1.09 \times 10^{1}$ | 0.916 | 0.913 |
| $\propto r^{-1}$ | 0.5 | $8.20 \times 10^{-2}$ | $3.46 \times 10^{-2}$ | $-6.10 \times 10^{2}$ | 1.0016 | 1.00056 | $-6.10 \times 10^{1}$ | 1.017 | 1.0056 |
| | 2 | $-3.18 \times 10^{-1}$ | $-2.70 \times 10^{-1}$ | $1.57 \times 10^{2}$ | 0.9937 | 0.9923 | $1.57 \times 10^{1}$ | 0.941 | 0.922 |
| $\propto r^{-7/4}$ | 0.5 | $2.57 \times 10^{-2}$ | $1.07 \times 10^{-2}$ | $-1.95 \times 10^{3}$ | 1.00051 | 1.00017 | $-1.95 \times 10^{2}$ | 1.0052 | 1.0017 |
| | 2 | $-1.05 \times 10^{-1}$ | $-1.06 \times 10^{-1}$ | $4.75 \times 10^{2}$ | 0.9979 | 0.9968 | $4.75 \times 10^{1}$ | 0.980 | 0.968 |

**Table 2.** The application of our results to the three disc density profiles, flat, $\propto r^{-1}$ and $\propto r^{-7/4}$. We use two disc masses and two disc radii. The encounter is coplanar and *retrograde* with $M_2/M_1 = 1$. Our initial energy and angular momentum take the values $E_{\rm initial} = 0$ and $J_{\rm initial} = 0.5$ respectively. $a'$ is the semi-major axis, $e'$ is the eccentricity and $r'_{\rm peri}$ is the periastron separation of the post-encounter binary orbit.

| $\Sigma(r)$ | $\frac{r_{\rm disc}}{r_{\rm peri}}$ | $\frac{\Delta E_{\rm orbit}}{M_{\rm disc}}$ | $\frac{\Delta J_{\rm orbit}}{M_{\rm disc}}$ | $\frac{M_{\rm disc}}{M_1} = 10^{-2}$: $\frac{a'}{r_{\rm peri}}$ | $e'$ | $\frac{r'_{\rm peri}}{r_{\rm peri}}$ | $\frac{M_{\rm disc}}{M_1} = 10^{-1}$: $\frac{a'}{r_{\rm peri}}$ | $e'$ | $\frac{r'_{\rm peri}}{r_{\rm peri}}$ |
|---|---|---|---|---|---|---|---|---|---|
| *const.* | 0.5 | $-2.86 \times 10^{-3}$ | $-8.17 \times 10^{-4}$ | $1.75 \times 10^{4}$ | 0.999943 | 0.999988 | $1.75 \times 10^{3}$ | 0.99943 | 0.999960 |
| | 2 | $-1.31 \times 10^{0}$ | $-6.35 \times 10^{-1}$ | $3.81 \times 10^{1}$ | 0.974 | 0.988 | $3.81 \times 10^{0}$ | 0.775 | 0.859 |
| $\propto r^{-1}$ | 0.5 | $-3.84 \times 10^{-3}$ | $-4.48 \times 10^{-4}$ | $1.30 \times 10^{4}$ | 0.999923 | 1.000020 | $1.30 \times 10^{3}$ | 0.99923 | 1.00020 |
| | 2 | $-9.10 \times 10^{-1}$ | $-4.13 \times 10^{-1}$ | $5.49 \times 10^{1}$ | 0.982 | 0.9925 | $5.49 \times 10^{0}$ | 0.833 | 0.919 |
| $\propto r^{-7/4}$ | 0.5 | $-8.04 \times 10^{-3}$ | $-1.26 \times 10^{-4}$ | $6.22 \times 10^{3}$ | 0.99984 | 1.000076 | $6.22 \times 10^{2}$ | 0.9984 | 1.00075 |
| | 2 | $-3.13 \times 10^{-1}$ | $-1.33 \times 10^{-1}$ | $1.60 \times 10^{2}$ | 0.9938 | 0.9978 | $1.60 \times 10^{1}$ | 0.939 | 0.977 |

The sign of the change in periastron of the orbit may be predicted from $\Delta E_{\rm orbit}$ and $\Delta J_{\rm orbit}$ using the standard equations of orbital mechanics to obtain

$$\Delta r_{\rm peri} = -\frac{J^3 \Delta E_{\rm orbit}}{4(GM)^2\mu^3} + \Delta J_{\rm orbit} = 0 \qquad (31)$$

or

$$\frac{\Delta J_{\rm orbit}}{\Delta E_{\rm orbit}} = \frac{J^3}{4(GM)^2\mu^3} \qquad (32)$$

$$= \frac{1}{4} \qquad (33)$$

for our parabolic orbit. For an increase in periastron, i.e. $\Delta r_{\rm peri} > 0$, then we have:

$$\frac{\Delta J_{\rm orbit}}{\Delta E_{\rm orbit}} \begin{cases} > \frac{1}{4} & \text{if } \Delta E_{\rm orbit} > 0 \\ < \frac{1}{4} & \text{if } \Delta E_{\rm orbit} < 0 \end{cases} ;$$

and for a decrease in $r_{\rm peri}$:

$$\frac{\Delta J_{\rm orbit}}{\Delta E_{\rm orbit}} \begin{cases} < \frac{1}{4} & \text{if } \Delta E_{\rm orbit} > 0 \\ > \frac{1}{4} & \text{if } \Delta E_{\rm orbit} < 0 \end{cases} .$$

Fig. 18 shows the ratio $\Delta J_{\rm orbit}/\Delta E_{\rm orbit}$ of the material at radius $r$ plotted against $r$ for both the prograde and retrograde encounters. In both cases the inner material affects the orbit in both senses. In the prograde case, material between $r/r_{\rm peri} = 0.6$ and $r/r_{\rm peri} = 1.8$ acts so as to decrease $r_{\rm peri}$, just as one might expect since the orbit is becoming more tightly bound. Outside $r/r_{\rm peri} = 1.8$, however, the material acts so as to increase $r_{\rm peri}$. This may be a little surprising but is due to the large increase in orbital angular momentum. In the retrograde case all material exterior to $r/r_{\rm peri} = 0.8$ acts so as to decrease $r_{\rm peri}$ as one might expect. We have not plotted $\Delta J_{\rm orbit}/\Delta E_{\rm orbit}$ at the smallest radii as we begin to suffer numerical error. It should be noted that in these regions (i.e. $r/r_{\rm peri} < 0.3$) linear theory becomes appropriate.

Manipulation of the standard orbital equations also yields:

$$\frac{\Delta E_{\rm orbit}}{E_{\rm orbit}} + 2\frac{\Delta J_{\rm orbit}}{J_{\rm orbit}} = -\frac{2e\Delta e}{(1-e^2)}. \qquad (34)$$

For initial orbits that are close to being parabolic, i.e. $e \sim 1$ or $E \sim 0$, then the $\Delta J_{\rm orbit}/J_{\rm orbit}$ term is small compared with the $\Delta E_{\rm orbit}/E_{\rm orbit}$ term and can be ignored, so that

$$\Delta e \propto \Delta E_{\rm orbit}. \qquad (35)$$

We can see this validated in Tables 1 and 2. All encounters with a positive energy change show an increase in eccentricity, while encounters that reduce the energy of the binary's orbit also reduce its eccentricity, the magnitude of the change in eccentricity being approximately proportional to the disc mass for a given disc density profile.

**Figure 18.** $\Delta J_{\rm orbit}/\Delta E_{\rm orbit}$ of the disc material at radius $r$ plotted against $r$ for the prograde (a – top) and retrograde (b – bottom) coplanar encounters. Points with $\Delta E_{\rm orbit} > 0$ are shown as filled squares whilst those with $\Delta E_{\rm orbit} < 0$ are shown as open squares.

## 6 CONCLUSIONS

We have used a reduced three-body method to evaluate the energy and angular momentum transfer between protostellar disc material and the orbit of a perturbing mass. Such interactions are expected to be relevant to the early stages of binary formation and in the dynamical evolution of compact stellar systems. We have concentrated on the response of a disc around a central star to the parabolic fly-by of a perturbing mass (without disc). Evidence for the non-coplanarity of discs within binary systems (Hale 1994) prompted us to investigate several non-coplanar encounter orientations as well as the coplanar retrograde and prograde encounters.

In almost all encounters we have found that both the energy and angular momentum transfers are dominated by disc material that becomes unbound from the system during the encounter. The only exception is in the case of the angular momentum transferred by material at large radius in the prograde encounter, here dominated by material remaining bound to the primary – the contributions from the unbound and captured material approximately cancelled each other out. We have found no correlation between the original locations of the material that yields peaks in energy transfer and that producing angular momentum peaks (Figs 4 and 11). This suggests that a different mechanism is responsible for the two effects. Temporal analysis of the peaks in the energy transfer showed that close encounters with the primary or secondary masses (or, occasionally, both) are responsible for peaks in energy transfer. Approximately one third of the energy transfer is in these peaks. Large transfers in angular momentum are caused by more distant encounters.

In all cases where there is some prograde element to the motion, we find a corotation resonance region where the disc material loses energy and angular momentum to the binary orbit (Fig. 9). The maximum extent of this region is out as far as $r/r_{\rm peri} = 0.6$ and thus for encounters more distant than this we expect the binary to become less strongly bound, and, if the original orbit were weak enough, the encounter could unbind the binary completely to produce two single stars. The resonance has two contributors. The inner part is caused by the disturbance of disc material that remains bound to the primary whilst the outer part is due almost entirely to material that is captured by the perturber. This region clearly cannot be treated in a linear fashion and the strength of the disturbance in the inner part implies that a linear treatment here may be doubtful. On the basis of this, we would expect linear calculations only to be valid out to a maximum radius of $0.25 r_{\rm peri}$ (i.e. valid only for encounters with $r_{\rm peri} \geq 4 r_{\rm disc}$) in the regime where $M_2 \sim M_1$. We therefore conclude that the assumption of Ostriker (1994), that linear theory could be extended to $r_{\rm peri} \sim 2 r_{\rm disc}$ for equal mass encounters, is overly optimistic. When $M_2 \ll M_1$ we would expect this linear regime to extend to larger radii owing to the weaker nature of the interaction, but cannot quantify this effect with the present (equal mass) calculations.

While CP93 reported that the prograde passage was most spatially destructive to the disc, this is not true energetically. We find that the retrograde encounter produces a larger energy and angular momentum transfer than the prograde case, by up to a factor of three or so. Also, the energy transferred during the encounter is greater than the binding energy of disc material exterior to periastron. The difference is dependent on the orientation of the encounter, from a factor of two in the prograde case to a factor of five for the retrograde encounter. This implies that previous work (Clarke & Pringle 1991a, for example), in which the energy transferred is set equal to the binding energy of the disc material exterior to periastron, is somewhat pessimistic, so that disc masses used in these calculations should be rescaled by a factor of between two and five.

As a first step in analysing this problem, we have ignored the gas dynamics of the disc material on the basis of its limited effect during the fly-by. It is clear from our results, however, that not all of the stripped disc material ends up on diverging orbits. Outwards of $r \sim 2r_{\rm peri}$, about half of the energy transferred is by material undergoing close encounters with the stars. Such material is back-scattered, forming converging flows (Figs 6 and 13) in which we would expect shock-heating of the gas to occur with subsequent energy loss. Obviously this cannot be quantified with the current calculation, but requires careful analysis in future hydrodynamic calculations. We note that such close, convergent encounters are considerably less important inward of $r \sim 2r_{\rm peri}$ and would therefore expect the current calculations to provide the best estimates in the limits of encounters that are not highly penetrative.

As is typical in such simulations, there is a multi-dimensional parameter space to be investigated, and such an exploration has to be an on-going project due to the massive amount of computing time required. It is our intention to begin this task, however, not only to establish some empirical relationships between the parameters that describe our models and the quantitative results, but also to determine the limits to which this reduced three-body approach may legitimately be applied before a more rigorous treatment is necessary.

Up to now, we have ignored the reaction of the binary orbit to the disc material during the encounter, only calculating a bulk change at the end of the interaction. It is necessary to perform some simulations taking this into account in order to evaluate how large the difference would be. Comparison of our post-encounter orbits with the parabolic initial orbit in the region near periastron where most of the interaction takes place suggests that any differences would be small. Viscous evolution of the disc material has also been ignored so far and an attempt to include this will also be made.

It has been emphasized previously that systems at this stage of their evolution are deeply embedded in clouds of dust and gas which severely limit our ability to see them. Violent interactions such as those demonstrated here are an efficient method of removing/disturbing a large fraction of this material, thus dramatically affecting both the photometry of the object and its subsequent evolution.